\documentclass{elsart}

\usepackage{epsfig}

\usepackage{amssymb}
\usepackage{graphicx}
\usepackage{floatflt}

\begin{document}

\begin{frontmatter}



\title{ Readout of GEM Detectors Using the Medipix2 CMOS Pixel Chip }


\author[label1]{A. Bamberger},
\author[label1,label3]{K. Desch},
\author[label1]{U. Renz},
\author[label1,label2]{M. Titov},
\author[label1]{N. Vlasov},
\author[label1,label3]{P. Wienemann},
\author[label1]{A. Zwerger}
\address[label1]{Albert-Ludwigs University of Freiburg, Physics Institute, Freiburg, Germany}
\address[label2]{Institute of Theoretical and Experimental Physics (ITEP), Moscow, Russia}
\thanks[label3]{now at Rheinische Friedrich-Wilhelms-University, Department of Physics, Bonn, Germany}


\begin{abstract}


We have operated a Medipix2 CMOS readout chip,
with amplifying, shaping and charge discriminating front-end electronics
integrated on the pixel-level,
as a highly segmented direct charge collecting anode in a three-stage 
gas electron multiplier (Triple-GEM) to
detect the ionization from $^{55}$Fe X-rays and electrons from 
$^{106}$Ru.
 The device allows to perform moderate energy spectroscopy measurements
(20 $\%$ FWHM at 5.9 keV $X$-rays) using only digital readout and two
discriminator thresholds.
 Being a truly 2D-detector, it allows to observe individual clusters
of minimum ionizing charged particles in $Ar/CO_2$ (70:30) and 
$He/CO_2$ (70:30) mixtures 
and to achieve excellent spatial resolution for position reconstruction 
of primary clusters down to $\sim 50~\mu m$,
based on the binary centroid determination method.


\end{abstract}

\end{frontmatter}


\section{Introduction}

 The development of Micro-Pattern Gas Detectors (MPGD), which has been initiated 
and still driven by elementary particle and nuclear physics, offers a great 
potential as a high resolution tracking detector for a variety of applications.
Recent advances in photolitography and microprocessing 
techniques from chip industry triggered the development
of more powerful detector concepts, such as Gas Electron Multiplier (GEM)~\cite{sauli1997} 
and the MICRO MEsh GAseous Structure (Micromegas)~\cite{giomataris1996}.
 The COMPASS fixed-target experiment at CERN has pioneered the use of large area
multi-GEM and Micromegas detectors for particle tracking at high intensities,
reaching 5~$kHz/mm^2$ close to the beam.
 Both technologies have achieved tracking efficiency close to 100~$\%$,
spatial resolution of the order of 70 - 100~$\mu m$
and time resolution of $\sim$~10~ns~\cite{gemcompass},~\cite{micromegascompass}.
 The excellent performance and radiation hardness of GEM and Micromegas 
detectors after several years of successful running in COMPASS has demonstrated 
the large-scale feasibility and robustness of the MPGD concept.
 GEM detectors have also entered the LHC project; they will be used 
in the LHCb Muon detector~\cite{lhcb} and in the TOTEM telescopes~\cite{totem}. 


 Recently, Micro-Pattern Gas Detectors were readout by high granularity
CMOS pixel chips with integrated amplification and digitization circuits (PixelASICs).
 This techniques opens novel detection possibilities for the application
of MPGDs in the future generation of particle and astrophysics experiments.
 A GEM detector coupled to a CMOS analog chip, comprising pixellated
charge collecting electrodes and readout electronics, can reconstruct the 
tracks of $^{55}$Fe photoelectrons with a length as short as
few hundred microns~\cite{bellazzini1}-~\cite{bellazzini4}. 
 Another possible active pixel anode plane for X-Ray Polarimetry 
is an amorphous silicon thin-film transistor (TFT) array, 
like those used in flat-panel monitors~\cite{black}.
 The use of pixellated gas detectors to enable true imaging of 
charged particle tracks has been also proposed
for an advanced Compton Telescope~\cite{bloser1}-\cite{takada}
and for the search of Weakly Interacting Massive Particles~\cite{sekiya}.

 The application of Micro-Pattern Gas Detectors for
high precision tracking at the future International Linear 
Collider (ILC) is an active field of $\rm R \& D$ in detector 
technology.
 A Time Projection Chamber (TPC) using MPGDs as a gas amplification device
is one of the main options for charged particle tracking~\cite{teslatdr},~\cite{ldc}. 
$R \& D$ is carried out within the ILC TPC collaboration aiming 
at the construction of a large prototype in the coming years. 
 While the standard approach to readout the signals is a segmented pad plane 
with front-end electronics attached through connectors from the backside,
an attractive possibility is the use of PixelASICs to serve as integrated 
device hosting the pad, the preamplification and the digitization 
and sparsification of the signals.
 This approach could offer an ultimate TPC resolution and a possibility
to observe individual electrons formed in the gas and count the number of
ionization clusters per unit track length for particle discrimination~\cite{hauschildtpc}.
 Earlier studies using GEM and Micromegas mounted on the Medipix2 chip provided
two-dimensional images of minimum ionizing track clusters~\cite{nikhef1}~-\cite{nikhef3}.

 In the following, we present results obtained with a Triple-GEM detector 
readout with Medipix2 ASIC, and irradiated with $^{55}$Fe X-rays and 
$^{106}$Ru electrons. An overview of the Triple-GEM~/~Medipix2 Detector,
built at Freiburg University, and of the Medipix2 readout and 
calibration system is given in Section~2.
 In Section~3.1, results obtained with the $^{55}$Fe source, including 
charge spectroscopy measurements using the dual threshold capability
of the Medipix2 digital readout, are described.
 In Section~3.2, results obtained with electrons from the $^{106}$Ru 
source are summarized. They include the measurement of the spatial 
resolution for primary ionization clusters in $Ar/CO_2$ and
$He/CO_2$ gas mixtures.
 Finally, long-term stability and operational experience with the 
present setup is discussed.

\section{The Triple-GEM~/~Medipix2 Detector}

\subsection{Experimental Setup}

The Triple-GEM/Medipix2 detector, built at Freiburg University, 
consists of three cascaded GEM multiplication stages coupled to a multi-pixel
ASIC developed at CERN (``Medipix2'')
with individual pixels that amplify, discriminate and count individual hits~\cite{medipix2},\cite{nimmedipix2}.
 The chip is designed and manufactured in a six-metal 0.25~$\mu m$ CMOS technology.
 Fig.~\ref{detectorsketch1} shows the schematic layout of the setup and
the enlarged photo of the Medipix2 ASIC.

\setlength{\unitlength}{1mm}
\begin{figure}[bth]
 \begin{picture}(70,70)
 \put(-5.0,-5.0){\includegraphics{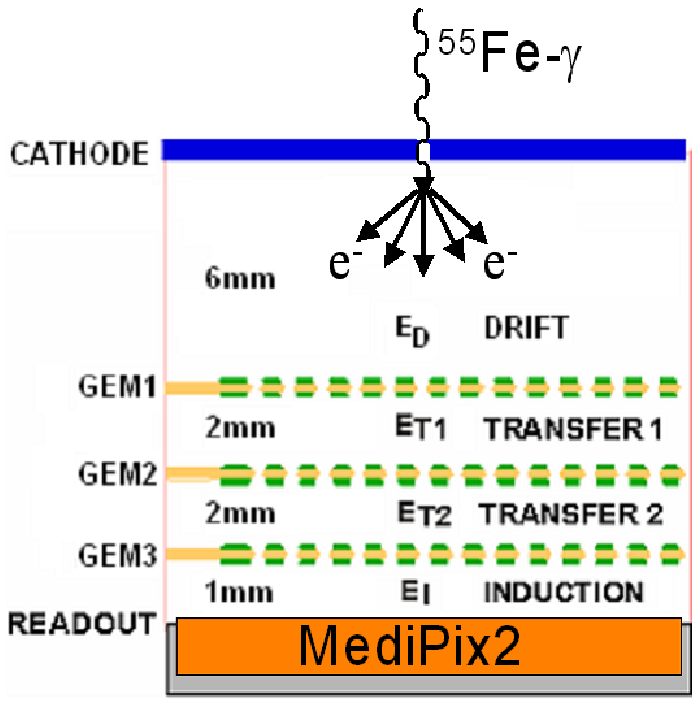}}
 \put(65.0, 0.0){\includegraphics{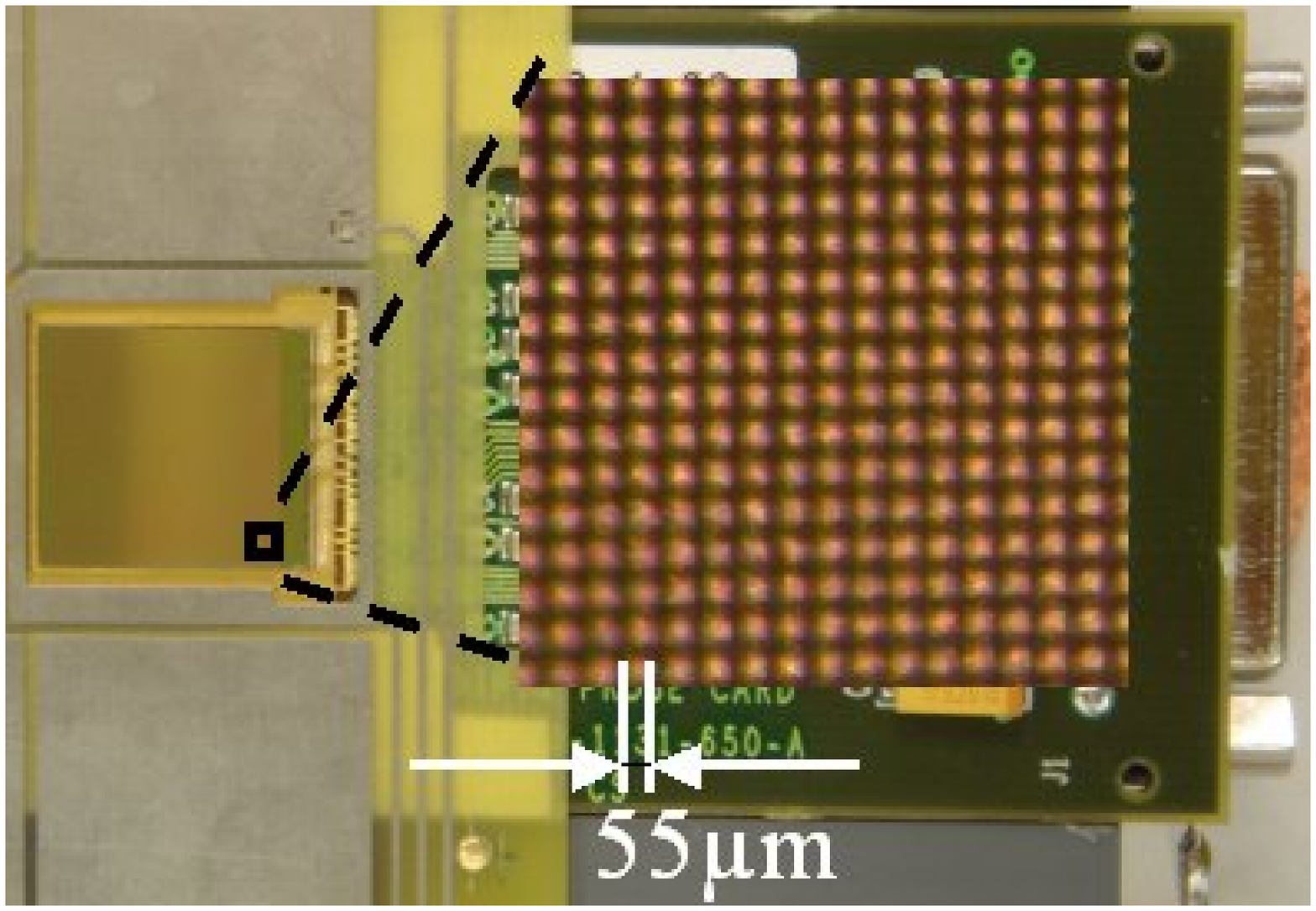}}
 \end{picture}
\caption[detectorsketch]
{(Left) A schematic drawing of the GEM/Medipix2 detector. $E_D$, $E_T$ and $E_I$
are the drift, transfer and induction field, respectively. (Right) Enlarged photo
of Medipix2 pixel cells; a 25~$\mu m$ wide conductive bump bond openings, 
used for electron collection, are seen as a matrix of dots.}
\label{detectorsketch1}
\end{figure}

\setlength{\unitlength}{1mm}
\begin{figure}[bth]
 \begin{picture}(70,70)
\put(15.0, -7.0){\includegraphics{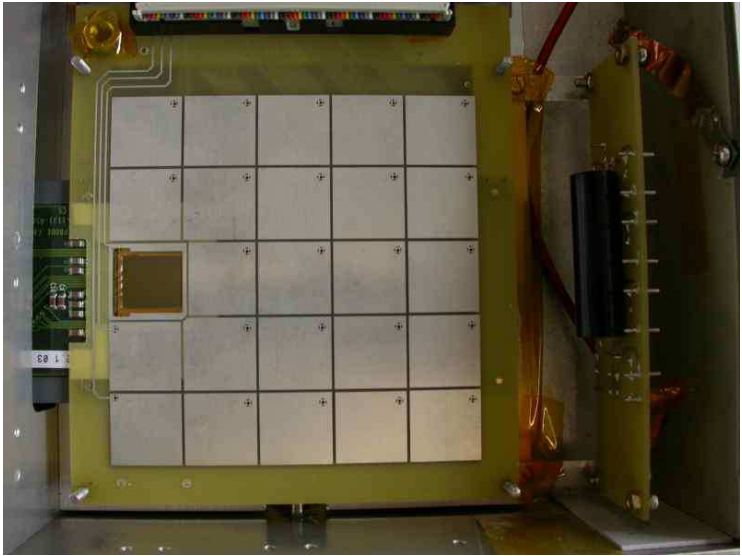}}
 \end{picture}
\caption[detectorsketch]
{Photo of the charge collection plane (anode) consisting of 25 pads, 
with one pad area replaced by the Medipix2 chip.}
\label{detectorsketch2}
\end{figure}

 The CERN-produced GEMs with double-conical holes, of 10~$\times$~10~$cm^2$ size,
have a standard thickness of 50~$\mu m$ with holes, 
arranged in a hexagonal pattern, of
140~$\mu m$ pitch and 70~$\mu m$ diameter (in Cu).
Application of a suitable voltage difference
between the metal layers of the GEM produces a strong electric field
in the holes ($\sim 50-100$~kV/cm), where the gas amplification occurs. 
 In multiple GEM structures there are 3 different electron drift regions:
in our setup, the drift gap (where the primary electrons are created) of 6 mm
has been chosen, while the transfer region (between successive GEMs) and 
the induction gap (between the last GEM and readout plane) are 2 and 1~mm,
respectively.
The 10~$\times$~10~$cm^2$ charge collection plane 
consists of 25 pads, each connected to a discrete preamplifier-discriminator
chain, developed for the L3 Forward Muon Detector~\cite{l3nima}.
Into one of the pads (approximately 1.8~$\times$~1.8~$cm^2$), 
a ``naked'' Medipix2 chip (without a silicon detector bonded)
is inserted (see Fig.~\ref{detectorsketch2}).
 The Medipix2 sensitive area is arranged as a
square matrix of 256~$\times$~256 pixels of 55~$\times$~55~${\mu m}^2$ size,
resulting in a detection area of 1.98~$cm^2$ which 
represents 87~$\%$ of the entire surface area.
 The periphery, placed at one side of the chip, includes 
the $I / O$ control logic, 13 8-bit DACs and 127 IO wire-bonding pads, arranged in a single 
row~\cite{medipix2},\cite{nimmedipix2}. The output of the Medipix2 preamplifier feeds
two identical discriminator branches (low and high thresholds), which can be set independently.
 Each pixel contains an 8bit configuration register. 
Six bits are used for the threshold equalization (three-bits for each discriminator),
one for masking noisy pixels and one to enable the test input pulse 
through the 8~fF-on-pixel capacitor.
 Using the serial or parallel readout interface, the readout of the whole pixel matrix
containing measured data takes 9~ms or 266~$\mu s$, respectively, for a 100~MHz clock.

 Approximately 75~$\%$ of each pixel is covered 
with an insulating passivation layer. Thus electrical field lines end on 
the conductive bump-bonding pads (octagonally shaped, 25 $\mu m$ wide)
exposed to the gas (see Fig.~\ref{detectorsketch1} (right)).
 Only electrons moving in the induction gap contribute to the signal.
The time development of the signal is fast as
the transit time of electrons in 1 mm induction gap is approximately 20~ns.
 The signal at the Medipix2 input is proportional to the charge,
which is collected on the bump-bonding pad in each pixel.
 This makes use of Medipix2 ASIC as a charge collecting anode and the pixel
segmented readout of a GEM detector, allowing a true 2D image reconstruction.

\subsection{Medipix2 data readout and calibration}
\label{calibration}

 The Medipix2 chip was controlled and read out by the MUROS2 
electronics~\cite{Medipix2readout1} and the software 
``Medisoft 4.0'' developed by University of Naples~\cite{Medipix2readout2}.
 Using a clock of 50 MHz in our setup, the pixel matrix was readout in
about 20~$ms$ in serial mode.
 To minimize the impact of threshold non-uniformity across the channels,
the optimization of the Medipix2 settings in Freiburg involved the
equalization of low ($THL$) and high ($THH$) thresholds 
by applying an external test pulse 
to the on-pixel 8~fF injection capacitance~\cite{nimmedipix2}.
 Using a pixel test input
a threshold scan for a fixed pulse charge
is performed to measure $S$-shaped curves for each channel:
from no pixel counts (0~$\%$ efficiency) to 100~$\%$ hits
for low threshold ($THL$) and, for larger input charges, 
from 100~$\%$ efficiency to no counts for the
high threshold ($THH$).
 The threshold dispersion between pixels is tuned based on the 50~$\%$-efficiency
point of the $S$-curve using the 3-bit-DAC available in each
discrimination branch.
 The difference in charge corresponding to 97.7~$\%$ and 2.3~$\%$ of the $S$-function,
divided by factor of four, is used as a measure of the equivalent noise charge (ENC) of the
pixel analog section, assuming a Gaussian distribution.

 A dedicated study was performed in order to obtain an absolute
calibration of the Medipix2 chip - 
to match low and high threshold $DAC$ settings ($THL$ and $THH$) 
to the corresponding effective charges in electrons ($q_{THL}$ and $q_{THH}$).
 In conventional Medipix setup the chip is connected to a Si semiconductor
detector allowing direct energy calibration with radioactive sources.
For our applications - we used a ``naked'' Medipix2 chip without $X$-ray
converter - calibration can only be performed using 
the electrical input pulse.
 In a first step, the resulting charge $Q^{test}_{input}$ (in $e^-$) 
appearing at the amplifier input can be estimated as a function
of applied voltage step $\Delta V$:
\begin{equation}
    Q^{test}_{input} = \frac{0.825 \cdot 8 fF}{1,6\cdot 10^{-19}} \cdot \Delta V,
    \label{testpuls}
\end{equation}
where $0.825$ is the amplification of the analog buffers used to transmit
the external test pulse to each pixel.

\setlength{\unitlength}{1mm}
\begin{figure}[bth]
 \begin{picture}(50,50)
 \put(0.0,-5.0){\includegraphics{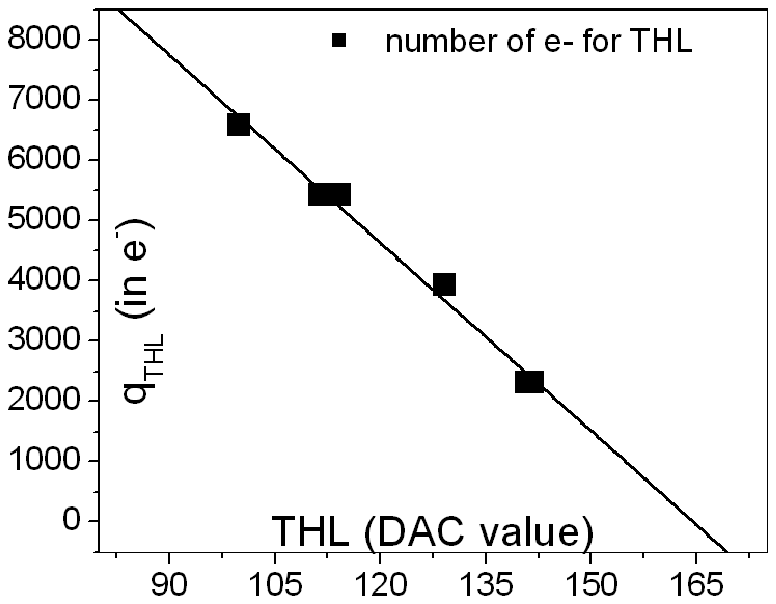}}
 \put(65.0,-5.0){\includegraphics{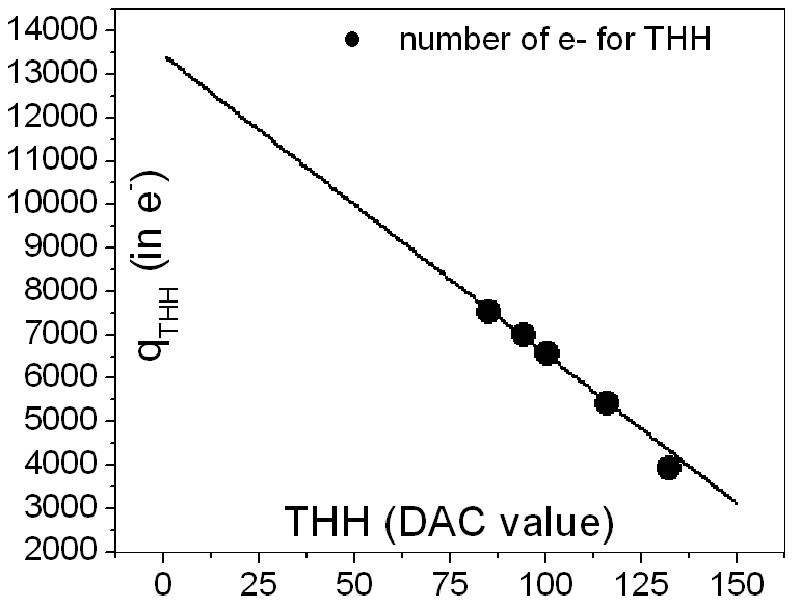}}
 \end{picture}
\caption[]{ Medipix2 calibration curves matching $DAC$ settings
for the low and high threshold values ($THL$ and $THH$)
to the effective charges in electrons ($q_{THL}$ and $q_{THH}$).}
\label{calibrationcurves}
\end{figure}

  Since it is known, that the behavior of the input buffers of the charge
injection circuit is non-linear for large input voltages ($>$ 100 mV)
we correct the obtained threshold charges $Q^{test}_{input}$ in Eq.~\ref{testpuls}
for the measured difference in response between various gamma sources and injected test pulses
for another Medipix2 ASIC, bonded to a $300 \mu m$ wafer of silicon~\cite{zwerger}. 
From this procedure, we obtain calibration curves 
for the $THL$ and $THH$ $DAC$ values shown in Fig.~\ref{calibrationcurves}.
 Applying the threshold equalization and using this calibration, 
we estimate the minimum operational low threshold $q_{THL}$
in our setup to be approximately 990~electrons with an
uncertainty of 140~electrons, resulting from the linear extrapolation
of the calibration curve in Fig.~\ref{calibrationcurves}.


\section{Measurements and Results}
\label{measurements}

 The performance of the Triple-GEM~/~Medipix2 detector was studied 
with $^{55}$Fe $X$-Rays and $^{106}$Ru electrons.
 The $^{55}$Fe 5.9~keV X-ray produces an ionization cluster in the detector volume,
corresponding to approximately 220 primary electrons in $Ar/CO_2$. 
The $^{106}$Ru source emits electrons
with a maximum kinetic energy of 3.54~MeV from the decay of $Rh^{106}$, 
which leaves the ionization track with 
approximately 60~(20) primary electrons per cm
in $Ar/CO_2$~($He/CO_2$) mixtures~(see Table~\ref{table1}).

\subsection{$^{55}$Fe X-rays in $Ar/CO_2$ (70:30)}

 The GEM/Medipix2 detector system,
was exposed to $^{55}$Fe $X$-rays, entering the
detector through the cathode drift electrode (see Fig.~\ref{detectorsketch1} (left)).
 Most of the $X$-rays are converted in the drift gap in $Ar/CO_2$~(70:30)
mixture emitting a photo-electron, which produces a short ionization track in the gas.
The cloud of primary electrons from the track drift through the 
multi-GEM structure, where they are multiplied and then collected on the
input pads of Medipix2 chip (see Fig.~\ref{detectorsketch1} (right)).
 Standard high voltage settings for $Ar/CO_2$~(70:30) operation were:
 drift field $E_D$~=~1.1~kV/cm, transfer field $E_T$~=~3.2~kV/cm, induction field
$E_I$~=~4.2~kV/cm, grounded anode readout plane and
$\Delta V_{GEM1}$ = $\Delta V_{GEM2}$ = $\Delta V_{GEM3}$~=~404~V,
corresponding to a gas gain of approximately $6 \times 10^4$ (see Fig.~\ref{gainestimate}). 

\setlength{\unitlength}{1mm}
\begin{figure}[bth]
 \begin{picture}(60,65)
 \put(-5.0,-5.0){\includegraphics{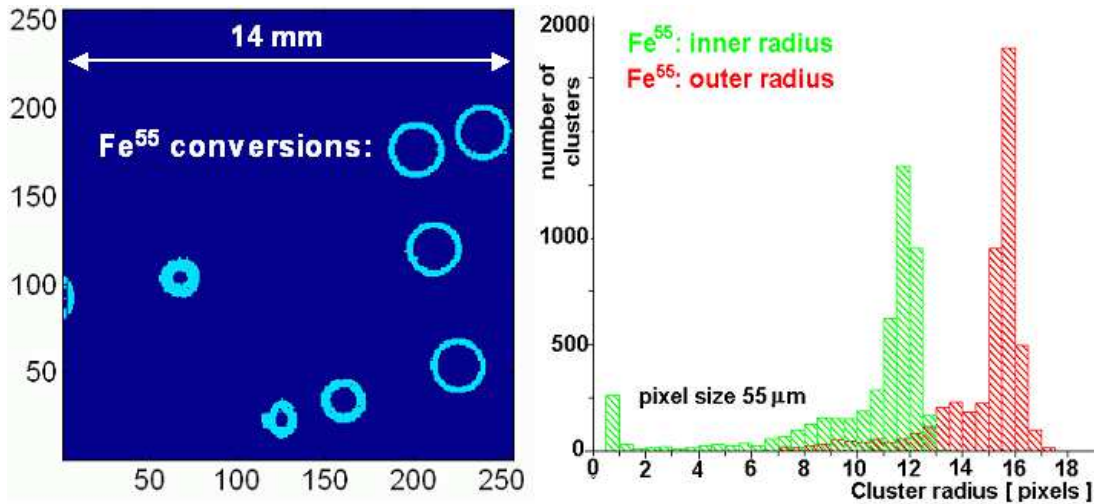}}
 \end{picture}
\caption{ (Left) Medipix2 images of $^{55}$Fe conversions. 
(Right) Distributions of inner and outer cluster radius of $^{55}$Fe ``donuts''
for chip operated in the charge window mode (see text).}
\label{fe55clusters}
\end{figure}

 The Medipix2 chip does not provide pulse height information, but it can
be operated in the charge window mode~\cite{medipix2}.
In this operation mode, the detected electron charge in each pixel is amplified and then
compared with Medipix2 low ($q_{THL}$) and high ($q_{THH}$) thresholds that form effectively 
a charge window $\Delta W = q_{THH} - q_{THL}$.
If the detected charge falls inside this window ($\Delta W$) a 13-bit digital counter 
is incremented. 
  With a digital readout in charge window mode, the $^{55}$Fe conversions
are seen as ``donuts'' of different sizes, according to the deposited energy. 
 Fig.~\ref{fe55clusters} (left) shows images of $^{55}$Fe quanta conversions, 
acquired without external trigger during approximately~1~s of Medipix2 acquisition time.
 To record this image the effective low and high thresholds were set to: 
$q_{THL} \approx$~$990~e^-$ (DAC $THL$ value~=~155) and 
$q_{THH} \approx 12000~e^-$ (DAC $THH$ value~=~20), 
respectively (see Fig.\ref{calibrationcurves}).
 A sample of $^{55}$Fe images was collected with these settings.
 Fig.~\ref{fe55clusters} (right) reveals clear peaks in the distributions
of inner ($r_{THH}$) and outer ($r_{THL}$) radius of nearly circular ``donuts'', 
which correspond to photoelectric conversions of 5.9 keV $X$-rays, with tails mostly 
coming from $Ar$-escape electrons and background events.

\setlength{\unitlength}{1mm}
\begin{figure}[bth]
 \begin{picture}(80,80)
 \put(35.0,-5.0){\includegraphics{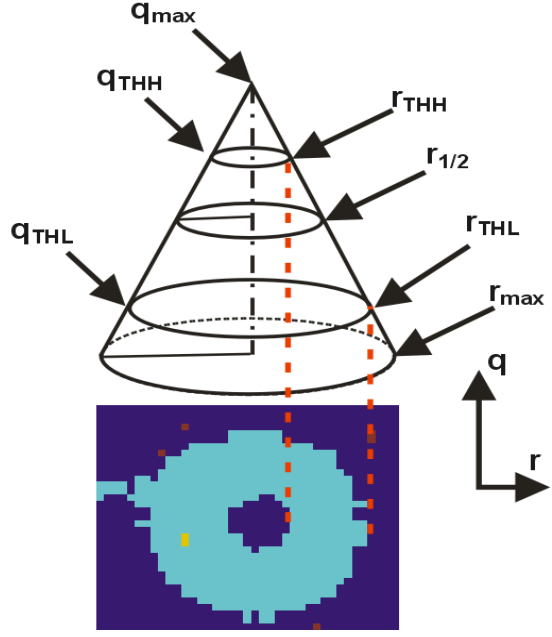}}
 \end{picture}
\caption{Schematics used for the calculation of the total cluster charge Q
from the inner ($r_{THH}$) and outer ($r_{THL}$) radii of the ``donut'' for the
low ($q_{THL}$) and high ($q_{THH}$) Medipix2
thresholds and assuming a conical shape of the charge cloud collected on the Medipix2
bump-bonding pads.}
\label{methodspectrum}
\end{figure}

\setlength{\unitlength}{1mm}
\begin{figure}[bth]
 \begin{picture}(65,65)
 \put(20.0,-7.0){\includegraphics{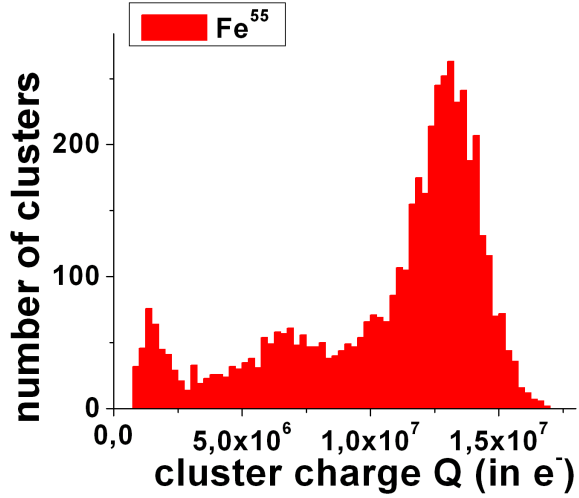}}
 \end{picture}
\caption{ Distribution of the cluster charge Q in ``donuts'' for the 5.9 keV $X$-rays, 
reconstructed using the method described in Fig.~\ref{methodspectrum}.}
\label{chargespectroscopy}
\end{figure}

\setlength{\unitlength}{1mm}
\begin{figure}[bth]
 \begin{picture}(65,65)
 \put(25.0,-7.0){\includegraphics{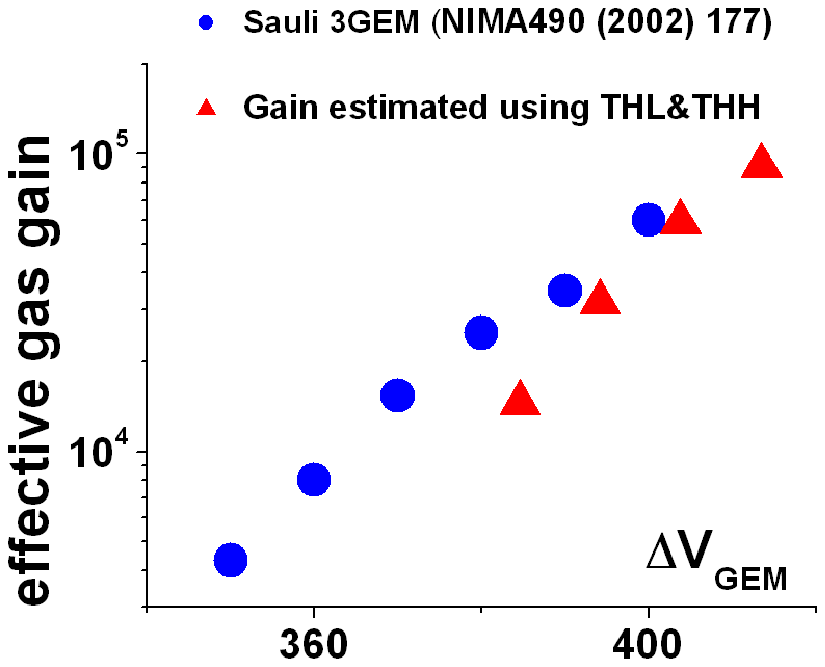}}
 \end{picture}
\caption{ Effective gas gain in $Ar/CO_2$~(70:30) 
calculated from the cluster charge Q in ``donuts'' (dots) and compared 
to the gas gain values, derived
from the measurement of the current for a known radiation flux (triangles) in~\cite{sauli4}.}
\label{gainestimate}
\end{figure}

   The operation of the Medipix2 ASIC in the charge window mode 
also allows us to perform an estimation of the total cluster charge 
in ``donuts''. The procedure is illustrated in Fig.~~\ref{methodspectrum}. 
 Assuming a conical shape of the charge cloud (the assumption of a more realistic
Gaussian profile would not alter the results significantly)
and using the inner ($r_{THH}$) and outer ($r_{THL}$) radius of the ``donut'' for the
corresponding settings of low ($q_{THL}$) and high ($q_{THH}$) 
thresholds, the total cluster charge Q can be estimated as:
\begin{equation}
     Q = \frac{1}{3} \pi \cdot r^2_{max} \cdot q_{max},
    \label{spectroscopy1}
\end{equation}
\begin{equation}
q_{max} = \frac{ r_{THL} \cdot q_{THH} -  r_{THH} \cdot q_{THL}}{r_{THL} - r_{THH}},
    \label{spectroscopy2}
\end{equation}
\begin{equation}
r_{max} = \frac{ r_{THL} \cdot q_{THH} -  r_{THH} \cdot q_{THL}}{q_{THH} - q_{THL}},
    \label{spectroscopy3}
\end{equation}
where $r$ is a dimensionless parameter, measured in number of pixels, $q$ is
a charge (in electrons).

 Fig.~\ref{chargespectroscopy} represents the distribution of 
the total cluster charge Q deposited by $^{55}$Fe 5.9~keV $X$-rays. 
The $Ar$-escape peak and photo-peak are clearly visible and separated.
The energy resolution of the ``charge spectroscopy'' method
is approximately 20$\%$ full-width at half-maximum (FWHM) for an 
$X$-ray energy of 5.9~keV. 
 By varying voltage across GEMs, the effective gas gain was determined from the 
central value of the photo-peak in the cluster charge distribution and 
the assumption that $^{55}$Fe quanta generate 220 primary electrons in $Ar/CO_2$.
 The results shown in Fig.~\ref{gainestimate}
are very similar to the absolute gain calibration, derived
from the measurement of the current for a known radiation flux in~\cite{sauli4}.
 

\subsection{ $^{106}$Ru electron tracks in $Ar/CO_2$~(70:30) and $He/CO_2$~(70:30)}
\label{ru106tracks}

 With the GEM/Medipix2 detector we collected a sample of 
tracks from a radioactive $^{106}$Ru $\beta^-$-source
for two different gas mixtures, $Ar/CO_2$~(70:30) and $He/CO_2$ (70:30).
 The gas gains were $6 \times 10^4$ ($\Delta V_{GEM}$~=404~V)
and $2 \times 10^5$ ($\Delta V_{GEM}$~=428~V) 
for the $Ar$ and $He$-based mixtures, respectively. 
 To record track images a four-fold coincidence of conventional readout pads
in a row, with a pixel chip positioned between them, 
was used to trigger the Medipix2 readout (see Fig.~\ref{coincidence_beta}).
 For these measurements, the chip was operated in a single discrimination mode 
(only low THL threshold was used).

\setlength{\unitlength}{1mm}
\begin{figure}[bth]
 \begin{picture}(65,65)
 \put(20.0,-5.0){\includegraphics{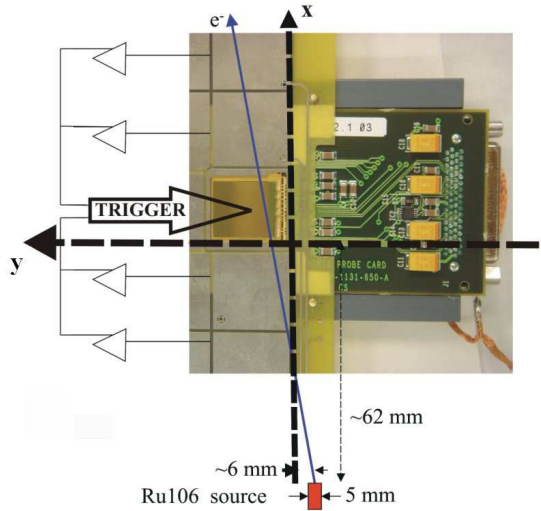}}
 \end{picture}
\caption[]
{Schematics of signal coincidence from four conventional readout pads used
to trigger Medipix2 readout. The relative position of $^{106}$Ru source with
respect to pixel ASIC in ($x$,$y$) plane is also indicated.}
\label{coincidence_beta}
\end{figure}

\setlength{\unitlength}{1mm}
\begin{figure}[bth]
 \begin{picture}(65,65)
 \put(-5.0,-5.0){\includegraphics{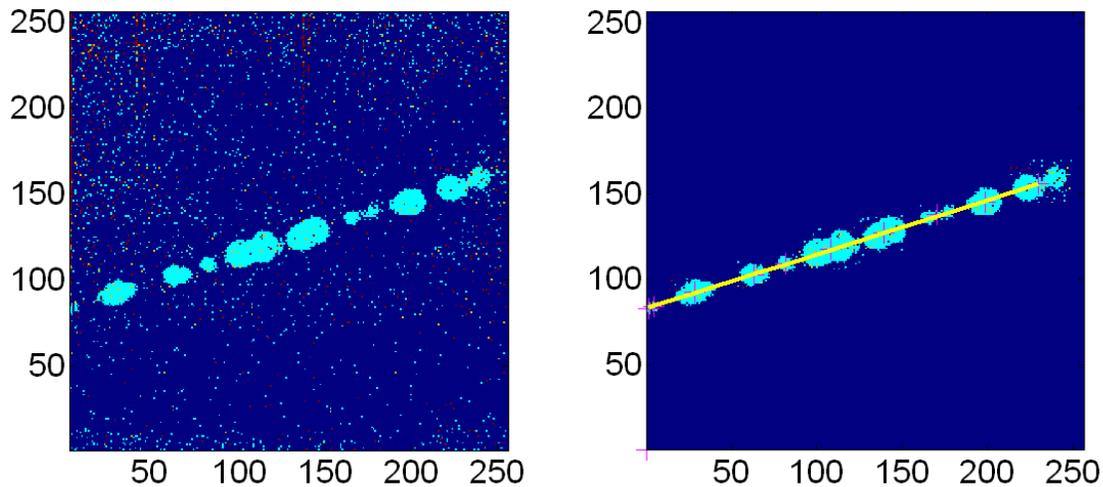}}
 \end{picture}
\caption{(Left) ``Raw'' Medipix2  image of the electron track from $^{106}$Ru source in
$Ar/CO_2$~(70:30). (Right) Straight line fit to the centers of clusters after the noise
suppression procedure (isolated noise hits, which are collected due to the relatively 
long recording time (up to 1~s), are suppressed).}
\label{tracksampleArCo2}
\end{figure}

\setlength{\unitlength}{1mm}
\begin{figure}[bth]
 \begin{picture}(60,60)
 \put(-5.0,-5.0){\includegraphics{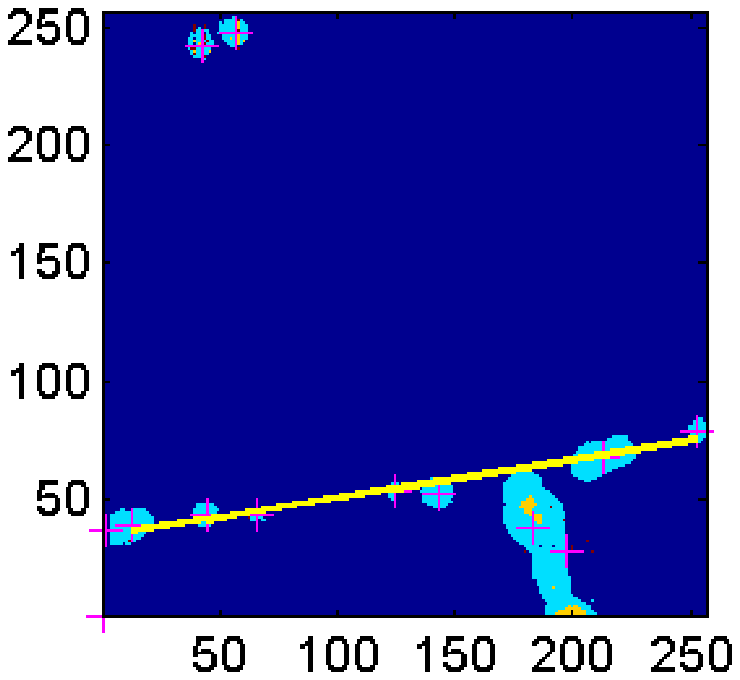}}
 \put(70.0,-5.0){\includegraphics{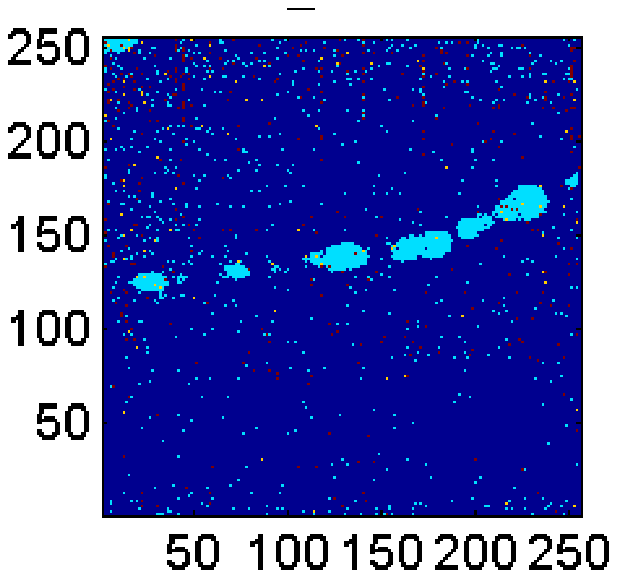}}
 \end{picture}
\caption{ Medipix2 image of $\delta$-electron liberated in 
$He/CO_2$ (70:30) with energy $E_{\delta} > 10~keV$ (left).
$^{106}$Ru electron track that suffered multiple Coulomb scattering in $Ar/CO_2$ (70:30) (right).}
\label{deltaray}
\end{figure}

 A recorded two-dimensional image of an electron track from $^{106}$Ru 
in $Ar/CO_2$ (70:30) is illustrated in Fig.\ref{tracksampleArCo2} (left)
along with a straight line fit to the reconstructed centers of clusters
in Fig.\ref{tracksampleArCo2} (right).
 The recorded electron track consists of clearly visible extended charge clusters. 
The observed cluster size varies considerably due to the fluctuating
number of primary ionization electrons per cluster 
and variations in the gas multiplication. 
 The average spatial extent of charge clouds on the Medipix2
surface depends on the sizes of the transfer and induction gaps 
and the electric fields strength $E_T$, $E_I$ and $\Delta V_{GEM}$.
The distribution of the $RMS$ spread of the cluster centers from a straight line
track is mainly determined by the electron diffusion in the drift region.
Fig.~\ref{deltaray} (left) shows an image of an energetic $\delta$-electron liberated
in $He/CO_2$. 
 Based on the visible $\delta$-electron range $R > 3.5$~mm,  
its energy $E_{\delta}$ can be estimated to be more than 10~keV.

\begin{table}[h]
\begin{center}
\begin{tabular}{|c|c|c|c|c|c|c|}\hline
 Gas mixture  &   $n_{p}^e$  & $n_{p}^\pi$ &  $n_{T}^e$ & $n_{T}^\pi$ & $X_{rad}$ $(m)$ &  $\rho$ ($g/cm^{3}$)\\\hline\hline
\hline
$Ar/CO_2~(70:30)$  &  28.9  & 30.3 & 62.8 & 76.4 & 125 & $1.8 \cdot 10^{-3}$  \\
$He/CO_2~(70:30)$  &  13.0  & 13.6 & 20.7 & 25.4 & 565 & $0.7 \cdot 10^{-3}$  \\ \hline
\end{tabular}
 \caption{ 
Number of primary ionization clusters $n_{p}^e$  
and total electron-ion pairs $n_{T}^{e}$ produced in 1~cm of 
$Ar/CO_2$ (70:30) and $He/CO_2$ (70:30) 
for 2~MeV electrons, as generated by 
HEED simulation program~\cite{hauschild}.
 The corresponding numbers ($n_{p}^\pi$,$n_{T}^{\pi}$) 
for minimum ionizing particles (e.g. 0.6~GeV pions) are presented 
for comparison.
 Radiation lengths $X_{rad}$ and gas mixture densities $\rho$ ($g/cm^{3})$
for both mixtures are also given.}
\label{table1}
\end{center}
\end{table}

 With an effective threshold of  $q_{THL} \approx 990 e^-$, on average 
$ N^{obs}_{cl} \approx$~8 clusters per track were reconstructed in both mixtures.
This number has to be compared with the expected
number of primary electron clusters $n_{p}^e$
released by a $\approx$~2~MeV $\beta^-$ track in a specific gas mixture (see Table~\ref{table1}).
  The determination of the efficiency for reconstructing a single or multi-electron
cluster, which requires a careful comparison with simulation, is under study.
 Here, we only give approximate estimate for the detection efficiencies.
 Based on simulations from the HEED program~\cite{hauschild}, 
$n_{p}^e \approx $ 40.5~(18.2) primary 
ionization clusters are expected along a track of 1.4~cm length in $Ar/CO_2$ ($He/CO_2$).
Thus, the average efficiency to reconstruct a primary 
ionization cluster is approximately 20~$\%$ (45~$\%$).
 Due to the fluctuations in the multiplication process in the GEMs we do not attempt
to estimate the single electron efficiency at this stage.
 However, it is obvious that the single electron efficiency is non-zero.
In particular, for $He/CO_2$ mixture only 3.2 multi-electron clusters are 
expected on average along a 1.4~cm track (see Table~\ref{table1a}), 
to be compared with $ N^{obs}_{cl} \approx$~8 recorded clusters.


\begin{table}[h]
\begin{center}
\begin{tabular}{|c|c|c|c|c|c|}\hline
 $k (e^-)$  &   1  &  2 &  3 & 4 & $\ge 5$  \\\hline\hline
\hline
$P(k)$ ($\%$) for $Ar/CO_2$  & 80.4  & 8.6 & 2.6 & 1.4 & 7.0   \\
\hline
$P(k)$ ($\%$) for $He/CO_2$  & 81.6  & 11.0 & 2.9 & 1.2 & 3.0  \\ \hline
\end{tabular}
 \caption{ 
 Cluster-size distribution probability $P(k)$ (in $\%$) 
of producing exactly $k$ ionization electrons 
for 2~MeV electrons in $Ar/CO_2$ (70:30) and $He/CO_2$ (70:30) 
mixtures, as generated by HEED simulation program~\cite{hauschild}.}
\label{table1a}
\end{center}
\end{table}

 In the following data analysis we focus on the achievable single point resolution
with the digital readout of the GEM/Medipix2 detector. 
 The following criteria were applied to select events, which were included in this analysis:
\begin{itemize}
\item A cluster is defined as a set of more than 4 adjacent hit pixels;
\item The distance between two neighbored track clusters
has to be smaller than 50 pixels (both in $x$ and $y$ directions);
\item More than 5 clusters per track are required;
\item Events with multiple tracks are rejected.
\end{itemize}

 A total of several hundred tracks were selected by this procedure for 
$Ar/CO_2$ and $He/CO_2$ gases.
 First, we perform a two-dimensional straight line fit to the cluster centers
and calculate the distance between each cluster center and the position of the 
point of closest approach along the fitted track. 
 The corresponding distributions, when all $N_{cl}$ cluster centers
are included in the track fit, are shown in Fig.~\ref{fitNclusters}.
 The standard deviations, $\sigma_{N}$, of a Gaussian fit to the residual distributions 
gives $64 \pm 2~\mu m$ for $Ar/CO_2$ (70:30) and $58 \pm 2~\mu m$ for $He/CO_2$ (70:30)
gas mixtures (see Table~\ref{resolutiontable}).
 In a second step, we repeat this procedure but omitting the cluster under
consideration from the track fit. 
  The corresponding distributions,  when the track fit is performed to $N_{cl}-1$ 
clusters, is shown in Fig.~\ref{fitN-1clusters}.
  The standard deviations, $\sigma_{N-1}$, are $84 \pm 3~\mu m$ and $73 \pm 3~\mu m$,
 respectively.
 An unbiased estimate for the single point resolution $\sigma_{mean}$ 
is obtained as the geometric mean of two methods~\cite{pointres}:
\begin{equation}
\sigma_{mean} = \sigma_{N} \cdot \sigma_{N-1},
\label{geometricmean}
\end{equation}
The corresponding resolutions are summarized in Table~\ref{resolutiontable}.

\setlength{\unitlength}{1mm}
\begin{figure}[bth]
 \begin{picture}(55,50)
 \put(0.0,-5.0){\includegraphics{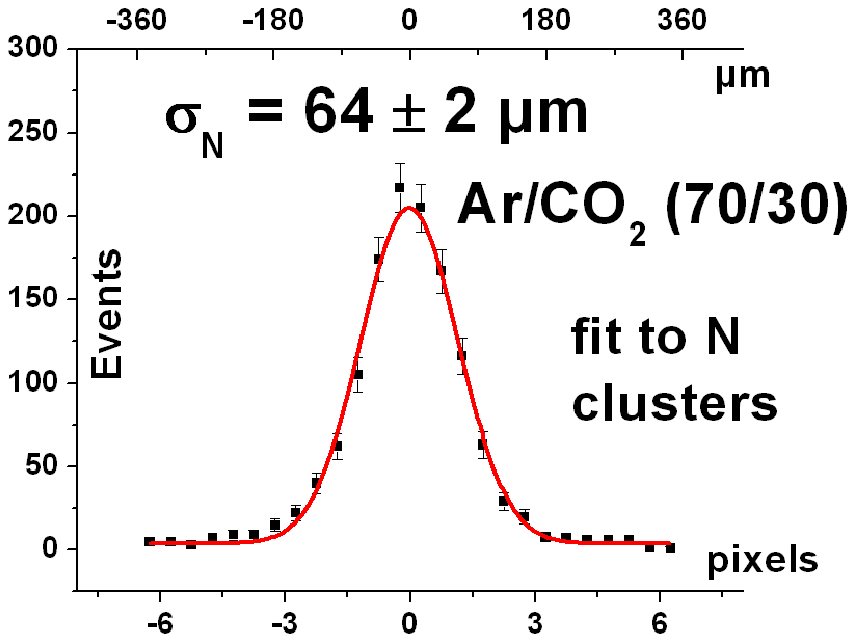}}
 \put(70.0,-5.0){\includegraphics{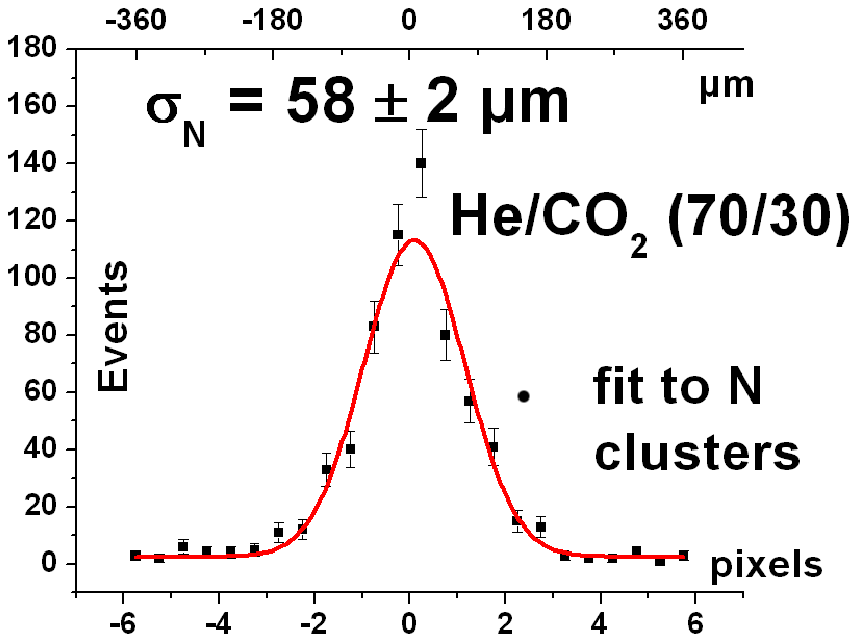}}
 \end{picture}
\caption{Single point resolution ($\sigma_N$) for $Ar/CO_2$ (70:30) (left)
and $He/CO_2$ (70:30) (right) when all $N_{cl}$ cluster centers
are included in the track fit. 
The residual distributions are well described by a single Gaussian.}
\label{fitNclusters}
\end{figure}

\setlength{\unitlength}{1mm}
\begin{figure}[bth]
\begin{picture}(55,50)
 \put(0.0,-5.0){\includegraphics{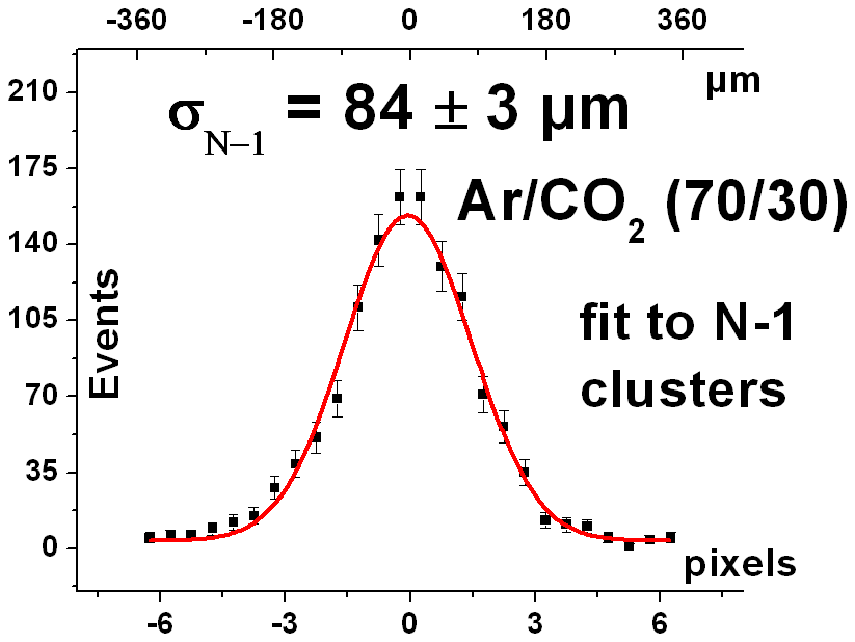}}
 \put(70.0,-5.0){\includegraphics{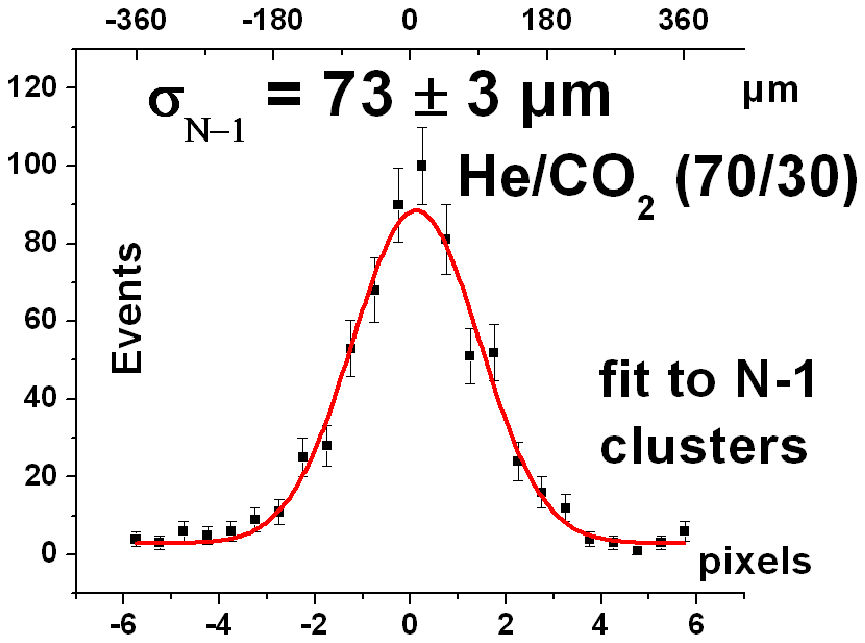}}
 \end{picture}
\caption{Single point resolution ($\sigma_{N-1}$) for $Ar/CO_2$ (70:30) (left)
and $He/CO_2$ (70:30) (right) when the track fit is performed to $N_{cl}-1$ 
clusters, excluding cluster under study.}
\label{fitN-1clusters}
\end{figure}

\setlength{\unitlength}{1mm}
\begin{figure}[bth]
 \begin{picture}(55,50)
 \put(0.0,-5.0){\includegraphics{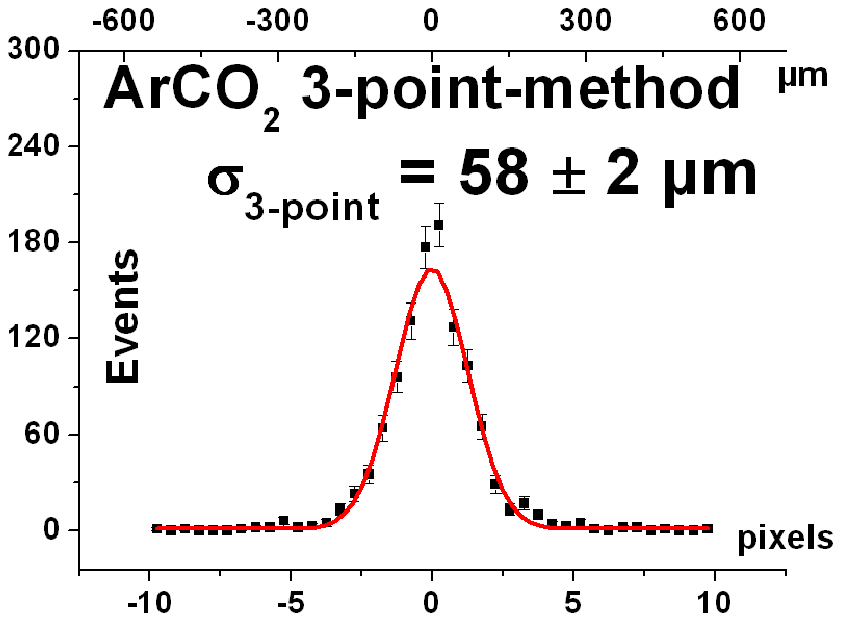}}
 \put(70.0,-5.0){\includegraphics{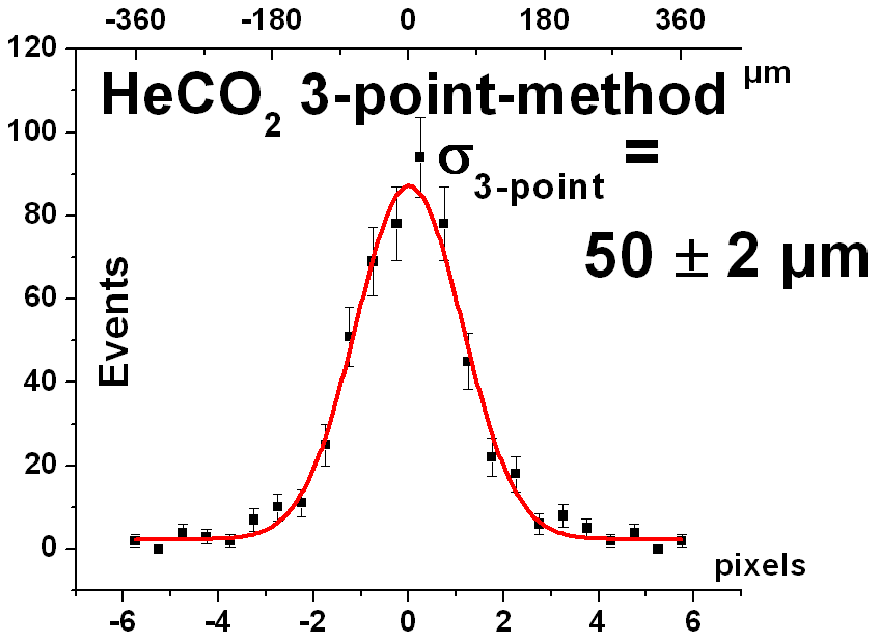}}
 \end{picture}
\caption{ The 3-point resolution ($\sigma_{3-point}$) for $Ar/CO_2$ (70:30) (left)
and $He/CO_2$ (70:30) (right).}
\label{3point}
\end{figure}

\begin{table}[h]
\begin{center}
\begin{tabular}{|c|c|c|c|c|c|}\hline
 Gas mixture         &   $\sigma_{N} (\mu m)$   & $\sigma_{N-1} (\mu m)$ & $\sigma_{mean} (\mu m)$ &  $\sigma_{mean}^{corr} (\mu m)$ &  $\sigma_{3-point} (\mu m)$ \\ \hline\hline
\hline
$Ar/CO_2 (70:30)$  &  $64 \pm 2$   & $84 \pm 3$ &  $73 \pm 3$ & $\sim 54$   &   $58 \pm 2$ \\
$He/CO_2 (70:30)$  &  $58 \pm 2$   & $73 \pm 3$ &  $65 \pm 3$ & $\sim 61$   &   $50 \pm 2$\\ \hline
\end{tabular}
 \caption{ Summary of resolution studies in $Ar/CO_2$ (70:30) and $He/CO_2$ (70:30)
mixtures using various evaluation methods: 
$\sigma_{mean}$ - unbiased spatial resolution, derived from geometric mean of two standard deviations
$\sigma_{N}$ and $\sigma_{N-1}$ (see Eq.~\ref{geometricmean} for details); $\sigma_{mean}^{corr}$ - 
single point resolution, 
determined from $\sigma_{mean}$ by correcting for multiple scattering effects, 
$\sigma_{3-point}$ - ``3-point'' resolution values. }
\label{resolutiontable}
\end{center}
\end{table}

\setlength{\unitlength}{1mm}
\begin{figure}[bth]
 \begin{picture}(60,60)
\put(20.0,-10.0){\includegraphics{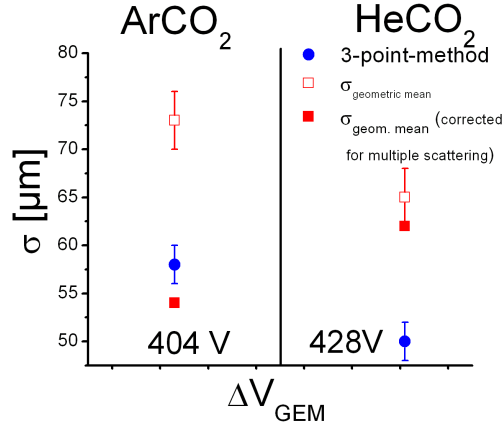}}
 \end{picture}
\caption{Summary of transverse single point resolution in $Ar/CO_2$ (70:30) and $He/CO_2$ (70:30)
mixtures using Medipix2 binary readout and centroid determination method.}
\label{resolutionplot}
\end{figure}

 In addition to inelastic collisions with atomic electrons, a few MeV $^{106}$Ru
electrons also suffer Coulomb scattering from nuclei, which increases the standard
deviations of residuals ($\sigma_N$ and $\sigma_{N-1}$) from the straight line fit. 
 This effect can be clearly seen from the image of a ``curved'' electron track 
in Fig.~\ref{deltaray} (right).
 We also calculated the ``3-point'' resolution, which is much less
sensitive to multiple scattering effects. 
For any three consecutive clusters ($i-1, i, i+1$) the ``3-point'' method calculates
the difference between the measured cluster $i$, 
and the position predicted from a straight line between cluster centers $i-1$ and $i+1$.
 The ``3-point'' resolution ($\sigma_{3-point}$) is obtained by multiplying
the $RMS$ of residuals by a factor of $\sqrt {2/3}$,
which yields an unbiased estimate of the single point resolution for the equally
spaced points.
 Here, the fit results are 71~$\mu m$ (61~$\mu m$) and the corresponding
``3-point'' resolutions ($\sigma_{3-point}$) are
58~$\mu m$ (50~$\mu m$) for the $Ar/CO_2$ and $He/CO_2$ mixtures, respectively
(see Fig.~\ref{3point}).
 The geometric mean method ($\sigma_{mean}$) gives a somewhat larger values
compared to the ``3-point'' values ($\sigma_{3-point}$) for both gases, 
owing to the non-negligible multiple scattering contribution of $\approx$~2~MeV electrons
to the achievable spatial accuracy.

 The effects of Coulomb scattering on the track parameters 
can be approximated by the appearance of the track curvature $c = 1 / R$
for the otherwise unscattered straight-line tracks.
 The corresponding multiple scattering $(\sigma_{mult~scat})$ contribution to the 
point measuring accuracy ($\sigma_{mean}$) can be evaluated 
from the variance of the electron trajectory $y = f(x) \sim x^2/2R$ over the track length of $L$:

\begin{equation}
    \sigma_{mult~scat} = <y^2> - <y>^2 = (\frac{1}{2 R})^2 \cdot (\frac{L}{2})^4 \cdot (\frac{1}{5} - \frac{1}{9})
\label{multscattering1}
\end{equation} 

 The variance $[c^2]$ of the total curvature, due to the multiple scattering, depends on 
the particle velocity $\beta$ and the momentum $p$ as well as on the radiation length $X_{rad}$
of the gas mixture~\cite{blum}:
\begin{equation}
    [c^2] = (\delta \frac{1}{R})^2 = (\frac{21~MeV}{\beta c p})^2 \cdot \frac{1}{X_{rad}} \cdot \frac{C_N}{2L}.
\label{multscattering2}
\end{equation} 
 The track length $L$ is measured in the Medipix2 plane ($L \approx 1.4~cm$) and
the constant $C_N$ is equal to $\approx \sqrt{2}$.
 By inserting ($1/R$) from Eq.~\ref{multscattering2} to the Eq.~\ref{multscattering1},
we may express the corresponding multiple scattering term as:
\begin{equation}
    \sigma_{mult~scat} = \frac{21~MeV}{\beta c p} \sqrt{ \frac{\sqrt 2}{128} \cdot \frac{L^3}{X_{rad}} \cdot (\frac{1}{5} - \frac{1}{9}).}
\label{multscattering3}
\end{equation} 

 Substituting the numerical values in Eq.~\ref{multscattering3} and assuming
the electron momentum of $p \approx 2~MeV$ one gets $\sigma_{mult~scat} 
\approx 49~\mu m~(23~\mu m)$ for the $Ar$ and $He$-based mixtures, respectively.
 The corrected values for the achievable single point accuracy,
obtained by quadratic subtraction of the multiple scattering contribution: 
$\sigma_{mean}^{corr} = \sqrt {\sigma^2_{mean} - \sigma^2_{mult~scat}}$,
are between 50~$\mu m$ and 60~$\mu m$, being consistent with ``3-point''
resolution results~(see Table~\ref{resolutiontable} and Fig.~\ref{resolutionplot})
for $Ar/CO_2$. For $He/CO_2$ small discrepancy still remains.

 Despite the large width of charge clouds ($\ge 10$ pixels)
in the GEM/Medipix2 readout plane (see Fig.~\ref{tracksampleArCo2}), 
 this result demonstrates the possibility to achieve spatial resolution
of the order of the pixel width,
based on the centroid determination method and digital Medipix2 readout.

\section{Long-Term Stability}

 A very attractive feature of the GEM is that it allows to completely decouple
charge amplifying region (GEM) from the collecting electrodes - 
printed circuit board, which operate at a unity gain.
 To date, no single Medipix2 chip, exposed to an induction field of $E_D \sim$~4~kV/cm, 
has been destroyed due to the electrostatic or GEM discharges 
in our triple GEM setup after several months of operation 
during the time-span of more than one year.
 No damage to a pixel VLSI analog chip used as a direct anode
of single GEM has been also observed in~\cite{bellazzini2}.
This proves that the reliable operation can be established for GEM
gas amplification coupled to $Si$-pixel readout and that the 
propagation of destructive discharges to the sensitive CMOS 
electronics is strongly suppressed even at large gas gains of $10^4 - 10^5$.

\section{Summary}


 The Medipix2 chip, being originally developed for a single photon counting, 
has been successfully adapted as a highly integrated pixel-segmented 
anode readout of a gas-filled detector, using a triple-GEM as a charge-multiplier.
 The CMOS readout concept offers the possibility of pixel sizes small enough
to observe individual primary electron clusters of minimum ionizing particle tracks 
and to provide real two-dimensional images of physics events.
 The approach holds a great potential for high-precision particle tracking 
at the next generation of high energy physics colliders
and for astrophysical applications.

 To evaluate the GEM/Medipix2 performance, we have carried out studies with
$^{55}$Fe and $^{106}$Ru sources.
 The Triple-GEM~/~Medipix2 detector allows to perform energy-sensitive 
charge spectroscopy measurements (20 $\%$ FWHM at 5.9 keV $X$-rays) 
using only digital readout and two discriminator thresholds.
 In the tracking applications, the detector has been shown to achieve an 
excellent spatial resolution of $\sim$~50~$\mu m$, based on the binary centroid
determination of the charge clouds, and allows to reconstruct tracks 
as short as $\sim$~1.5~$cm$ length.


 For the ILC applications the use of CMOS pixel ASIC in the TPC readout plane
will allow to observe individual electron clusters in 3D and
fully exploit unprecedented 3D-granularity for the 
gaseous tracking and to minimize the material in the endplate, 
since front-end electronics is naturally integrated into the
readout pad plane.
 A modification of the Medipix2 chip (``TimePix'') to measure also the
drift time information of primary electrons is under development~\cite{timepixproposal},\cite{eudet}.
 This will enable to measure not only the 2D projection, but also to reconstruct 
the true 3D-space points of charged particles crossing a large TPC volume.
 The possibility to use time-over-threshold information for time-walk
correction and charge estimation are also foreseen in the chip design.

\section*{Acknowledgments.}

 We thank the Medipix Collaboration for providing us with several Medipix2 chips,
and for the readout software and hardware. We would like to thank to Michael
Campbell, Erik Heijne, Xavier Llopart and Fabio Sauli 
for stimulating discussions and a lot of valuable advices.

\end{document}